\documentstyle[12pt]{article} 
\begin{document}
\baselineskip=20pt

\begin{center}
{\Large {\bf Normal ordering and boundary conditions in open bosonic strings }}\\
\vspace{1cm} 

{\large Nelson R. F. Braga$^{a}$ , Hector L. Carrion$^{a}$ and 
Cresus F. L. Godinho$^{b}$  } \\
 
\vspace{0.5cm}
{\sl
$^a$ Instituto de F\'{\i}sica, Universidade Federal
do Rio de Janeiro,\\
Caixa Postal 68528, 21941-972  Rio de Janeiro, Brazil\\[1.5ex]

$^b$ Centro Brasileiro de Pesquisas F\'{\i}sicas, Rua Dr Xavier Sigaud 150,\\
 22290-180,
Rio de Janeiro RJ Brazil }
\end{center}
\vspace{1cm}

\abstract  
Boundary conditions play a non trivial role in string theory. 
For instance the rich structure of D-branes is generated
by choosing appropriate combinations of Dirichlet and Neumann boundary 
conditions. Furthermore,  when an antisymmetric background is present 
at the string end-points (corresponding to mixed boundary conditions) 
space time becomes non-commutative there. 

We show here how to build up normal ordered products for bosonic string position 
operators that satisfy both equations of motion and open string boundary conditions
at quantum level.
We also calculate the equal time commutator of these normal ordered products
in the presence of antisymmetric tensor background.


\vspace{1cm}

\vspace{1cm}

\noindent braga@if.ufrj.br; godinho@cbpf.br ; mlm@if.ufrj.br

\vfill\eject

\section{Introduction}

Recent progress in string theory\cite{Malda} indicates a scenario where our 
four dimensional space-time should correspond to a D-brane\cite{Po} 
representing the boundary of a larger manifold. This idea also 
proved useful indicating a possible explanation for the
hierarchy problem\cite{RS1,RS2}. 
One important consequence of such a model is the non-commutativity of space 
time coordinates in our four dimensional world\cite{CH1,Alekseev:1999bs,SW}.
The reason is that D-branes correspond to the space where open string 
endpoints are located and where the corresponding string boundary conditions must 
be satisfied. In the presence of an antisymmetric tensor background 
these conditions are incompatible with commuting coordinates. 

Since antisymmetric fields show up in the massless spectrum of closed 
strings living on the D-branes it is reasonable to suspect that our physical world
could be non-commutative at very small length scales.
This is one of the reasons for the increasing interest
in studying many aspects of non-commutative quantum field theories
as can be seen for example in\cite{SW,REVIEW}.   
Furthermore, this fact illustrates the non trivial role of boundary 
conditions in string theory and the importance of taking them 
into account when considering the quantization of open strings.

In quantum field theory, products of quantum fields at the same space-time points 
are in general singular objects. The same thing happens in string theory if one 
multiplies position operators, that can be taken as conformal fields on the world sheet.
This situation is well known and one can remove the singular part of the 
operator products by defining normal ordered well behaved objects\cite{Po2}.
This is important, for example,  when one builds up the generators of conformal 
transformations and investigate the realization, at quantum level,
of the classical symmetries.

Normal ordered products of operators are usually defined so as to satisfy
the classical equations of motion at quantum level.
The purpose of this article is to define normal ordered products for open string position
operators that additionally satisfy the boundary conditions.
This way we will define a normal ordering that will be valid also at string end-points.
We will also investigate the relation between this new definition for normal ordering
and the non commutativity of space time coordinates.

\section{String position operator products }
 
The classical action for a bosonic string in the presence of a constant 
antisymmetric background taking a world sheet with Euclidean signature is 

\begin{equation}
\label{1}
S \,=\,
{ 1 \over 4\pi \alpha^\prime }
\int_{\Sigma} d^2\sigma \Big(
g^{ab} \eta_{\mu\nu } \partial_a X^\mu \partial_b X^\nu \,- \,i\, 
\epsilon^{ab} B_{\mu\nu} \partial_a X^\mu \partial_b X^\nu 
 \Big)
\end{equation}

\noindent where $X^\mu $ are the spacetime string coordinates and $ B_{\mu\nu}$
is the antisymmetric field. The string world sheet $\Sigma$ is represented by the
parameters $\sigma_1 \equiv \tau \,,\,\sigma_2\equiv \sigma\,$
with,  as usual, the boundary (string endpoints) at $\sigma = 0\,,\,\pi\,$. 
The Euclidean world sheet metric is $g^{\tau\tau} = g^{\sigma\sigma} =1 $ 
and the antisymmetric tensor is chosen by  $\epsilon^{\tau\sigma} = 1$. 

The variation of the action gives us 
a volume term that vanishes imposing the equations of motion

\begin{equation}
(\, \partial^2_\tau \, +\, \partial^2_\sigma\,)  X^\mu\,=\,0
\end{equation}

\noindent plus a boundary term that vanishes if we additionally impose  
that the string coordinates satisfy the boundary conditions 

\begin{eqnarray}
\label{e1}
\Big( \eta_{\mu\nu } \partial_\sigma X^\nu &+& i
 B_{\mu\nu} \partial_\tau X^\nu \Big)\vert_{_{\sigma = 0}}  \,=\,0\nonumber\\
\Big( \eta_{\mu\nu } \partial_\sigma X^\nu &+& i  B_{\mu\nu} \partial_\tau X^\nu 
\Big)\vert_{_{\sigma = \pi}}
\,=\,0
\end{eqnarray}

This boundary conditions, when imposed at quantum level, are responsible for the
non commutativity of the position operators\cite{CH1,SW}. We can infer this result
by realizing that this conditions represent constraints in phase space 
relating position and conjugate momenta. 

It is convenient, for studying the quantum operators,
to introduce complex world sheet coordinates: 
$ z \,=\,  \tau + \imath \sigma \,$, $\,\bar{z} \,=\, \tau - \imath \sigma \,$;
$ \partial_{z}\, = \, 1/2 ( \partial_{\tau} - \imath \partial_{\sigma})\, $,
$\, \partial_{\bar{z}} \,= \, 1/2 ( \partial_{\tau} + \imath \partial_{\sigma} )\,$

The action takes the form

\begin{equation}
\label{CA}
S \\\,=\,\frac{1}{2 \pi \alpha} \int dz^2 \, [\, \eta_{\mu\nu }\partial_z X^{\mu}
\partial_{\bar z} X^{\nu} -  B_{\mu\nu} \partial_z X^{\mu}
\partial_{\bar z} X^{\nu} \, ] \label{acao1}
\end{equation}

\noindent while classical equations of motion and boundary conditions take the form
\begin{equation}
\partial_{\bar z} \partial_z  X^i \,=\, 0
\end{equation}
 
\begin{eqnarray}
\Big( \eta_{\mu\nu} ( \partial_z - \partial_{\bar z} ) + 
 B_{\mu\nu} ( \partial_z + \partial_{\bar z} ) \Big) X^\nu\vert_{_{z = \bar z}} &=& 0
\nonumber\\
\Big( \eta_{\mu\nu} ( \partial_z - \partial_{\bar z} ) + 
 B_{\mu\nu} ( \partial_z + \partial_{\bar z} ) 
\Big) X^\nu\vert_{_{z = \bar z + 2\pi\,i}} &=& 0
\end{eqnarray}

We can study the properties of quantum operators by considering the 
expectation values of the corresponding classical objects. Defining 
the expectation value of an operators ${\cal F}$ as\cite{Po2}

\begin{equation}
\langle {\cal F} [X] \rangle \,=\,\int [dX] exp ( - S[X] ) {\cal F} [X] 
\end{equation}

\noindent and using the fact that the path integral of a total derivative vanishes one
finds that the equations of motion and boundary conditions are 
realized for the expectation values of string coordinates $X^\nu$ 

\begin{eqnarray}
0 &=& \int [dX] {\delta \over \delta X^\nu (z^\prime , {\bar z}^\prime )}
exp ( - S[X] ) \,=\,
\Big\langle \,{1\over \pi\alpha^\prime } \partial_{\bar z^\prime}
\partial_{z^\prime} X_\nu ( z^\prime, \bar z^\prime ) \Big\rangle \nonumber\\
&+& {1\over 2\pi \alpha^\prime}  \oint_{_{\partial \Sigma}} \delta^2 (z - z^\prime) 
\Big\langle \Big( \eta_{\nu\mu} ( \partial_z - \partial_{\bar z} ) + 
 B_{\nu\mu} ( \partial_z + \partial_{\bar z} ) \Big) X^\mu (z, {\bar z} ) dz
 \,\Big\rangle \, = \,0\,\,.
\end{eqnarray}

\noindent The last (singular) term is integrated over the boundary, 
where $dz = d \bar z$. This equation implies that both string equations of motion 
and boundary condition hold as expectation values.
So the corresponding quantum position operators satisfy the equivalent 
conditions (as long as they are not multiplied by other operators located at 
the same world sheet point)

\begin{equation}
\partial_{\bar z} \partial_{z} {\hat X}^\nu ( z, \bar z ) = \,0
\end{equation}

\begin{eqnarray}
\Big( \eta_{\nu\mu} ( \partial_z - \partial_{\bar z} ) + 
 B_{\nu\mu} ( \partial_z + \partial_{\bar z} ) \Big) {\hat X}^{\mu}\vert_{_{z = \bar z}} 
&=& 0 \nonumber\\
\Big( \eta_{\nu\mu} ( \partial_z - \partial_{\bar z} ) + 
 B_{\nu\mu} ( \partial_z + \partial_{\bar z} ) \Big) 
{\hat X}^{\mu}\vert_{_{z = \bar z + 2\pi\,i}} &=& 0
\end{eqnarray}

Products of operators at the same point will have a singular behavior.
We can see this by calculating

\begin{eqnarray}
0 &=& \int [dX] {\delta \over \delta X^\nu (z^\prime , {\bar z}^\prime )}
exp ( - S[X] ) \,\, X^\rho (z^{\prime \prime} , {\bar z}^{\prime\prime} ) 
\nonumber\\ &=&
\Big\langle \,\delta^2 ( z^{\prime} - z^{\prime \prime} )\delta^{\,\,\rho}_\nu
\,+\, ({1\over \pi\alpha^\prime } \partial_{\bar z^\prime}
\partial_{z^\prime} X_\nu ( z^\prime, \bar z^\prime ) 
X^\rho (z^{\prime \prime} , {\bar z}^{\prime\prime} ) \nonumber\\
&+& {1\over 2\pi \alpha^\prime}  \oint_{_{\partial \Sigma}} \delta^2 (z - z^\prime) 
\Big( \eta_{\nu\mu} ( \partial_z - \partial_{\bar z} ) + 
 B_{\nu\mu} ( \partial_z + \partial_{\bar z} ) \Big) X^\mu ( z, \bar z) 
X^\rho (z^{\prime \prime} , {\bar z}^{\prime\prime} )  dz
 \,\Big\rangle \, = \,0\,\,.\nonumber\\
& &
\end{eqnarray}

\noindent The volume term gives an extra singular term to the equation of motion
for a product of two fields

\begin{equation}
{1\over \pi \alpha^\prime } \, \Big\langle \, 
\partial_{z^\prime} \partial_{\bar z^\prime} \, 
X^\mu (z^\prime, {\bar z}^\prime) X^\nu (z^{\prime\prime}, {\bar z^{\prime\prime}}  )\,
\Big\rangle \, = \,- \eta^{\mu\nu} \,\Big\langle \, \delta^2
( z^\prime - z^{\prime\prime} , {\bar z}^\prime - {\bar z}^{\prime\prime} ) \, 
\Big\rangle 
\end{equation}

\noindent while the boundary terms vanishes if this product of two fields satisfies 
the same boundary condition as the single field

\begin{equation}
\Big\langle \Big( \eta_{\nu\mu} ( \partial_{z^\prime} - \partial_{{\bar z}^\prime} ) + 
 B_{\nu\mu} ( \partial_{z^\prime} + \partial_{{\bar z}^\prime}  ) \Big) 
 X^\mu  (z^\prime, {{\bar z}^\prime} ) X^\rho (z^{\prime\prime}, 
{\bar z^{\prime\prime}} )  \vert_{_{Bound.}} \,\Big\rangle \,=\, 0\,,
\end{equation}
 
\noindent where $Bound.$ means that we are taking this condition both at 
$z = \bar z $ and at $ z = \bar z + 2\pi\,i\,$.  
Thus the products of operators will satisfy 

\begin{equation}
\partial_{\bar z^\prime} \partial_{z^\prime} {\hat X}^\mu ( z^\prime, {\bar z}^\prime) 
{\hat X} ^\nu (z^{\prime\prime}, {\bar z^{\prime\prime}} ) = \,- \pi \alpha^\prime\,
\eta^{\mu\nu} 
\, \delta^2
( z^\prime - z^{\prime\prime} , {\bar z}^\prime - {\bar z}^{\prime\prime} )
\end{equation}

\begin{equation}
\Big( \eta_{\nu\mu} ( \partial_{z^\prime} - \partial_{\bar z^\prime} ) + 
 B_{\nu\mu} ( \partial_{z^\prime} + \partial_{\bar z^\prime} ) \Big) 
{\hat X}^\mu ( z^\prime, {\bar z}^\prime)
{\hat X}^\rho ( z^{\prime\prime}, {\bar z^{\prime\prime}} )
\vert_{_{Bound.}} \,=\, 0
\end{equation}

If we define a normal ordered product of two position operators in the standard way 
\cite{Po2}

\begin{equation}
:\,{\hat X}^\mu (z, \bar z )\, {\hat X}^\nu (z^\prime, {\bar z}^\prime)\,:
\,=\, {\hat X}^\mu (z , \bar z  ) \,{\hat X}^\nu (z^\prime ,{\bar z}^\prime)\, + 
{ \alpha^\prime \over 2} \eta^{\mu\nu} ln \vert z - z^\prime \vert^2
\end{equation} 
  
\noindent it satisfies the equation of motion at quantum level:

\begin{equation}
\partial_{\bar z} \partial_{z} 
:\,{\hat X}^\mu ( z, \bar z )\, {\hat X}^\nu (z^\prime ,{\bar z}^\prime )\,:\,=\,0
\end{equation}

\noindent but fails to satisfy the boundary conditions.  So we will introduce
a different kind of normal ordered product satisfying both equation of motion 
and boundary conditions.

The mathematical problem posed by defining the normal ordering is related  to that 
of calculating Green's functions\cite{AGNY,Schomerus:2002dc,DN,Dolan:2002px}. 
The normal ordered product is defined by subtracting out the corresponding
Green's functions. 
So we can find normal ordered products satisfying open string boundary condition 
using the solutions to open string Green's functions.

At this point it is more convenient to choose world sheet coordinates that simplify the
representation of the boundary. In the present coordinates the 
boundary $\sigma = 0\,$ corresponds to $z = {\bar z} ,$ and $\sigma = \pi$ 
to $\,z = {\bar z} + 2\pi\,i $. Introducing 

$$ w = e^{\tau + i \sigma} \,\,\,;\,\,\,{\bar w} = e^{\tau - i\sigma} $$

\noindent the complete boundary corresponds just to the region $w = {\bar w}$.
On the other hand the factor $w {\bar w}$ in  $ \partial_z \partial_{\bar z} \,=
\, w {\bar w}\partial_w \partial_{\bar w} \,$ cancel out precisely the Jacobian
of the coordinate transformation in such a way that the action in terms of 
$w, {\bar w}$ has still the same form as in eq. (\ref{CA}). The boundary conditions 
take the form

\begin{eqnarray}
\Big( \eta_{\mu\nu} ( w\partial_w &-& {\bar w}\partial_{\bar w} ) + 
 B_{\mu\nu} ( w\partial_w + {\bar w}\partial_{\bar w} ) \Big) 
{\hat X}^\nu\vert_{_{w = \bar w}}\nonumber\\ &=& 
 w\,\Big( \eta_{\mu\nu} ( \partial_w - \partial_{\bar w} ) + 
 B_{\mu\nu} ( \partial_w + \partial_{\bar w} ) \Big) 
{\hat X}^\nu\vert_{_{w = \bar w}} \,=\, 0\,\,.
\end{eqnarray}

This implies that starting with a solution in coordinates 
$z, {\bar z}$ that satisfies the boundary conditions just at $\sigma = 0$ 
and replacing everywhere  $\,z ,{\bar z}$ by   $ w ,{\bar w} $ we get a 
new solution that satisfies the boundary conditions both at $\sigma = 0$ and 
$\sigma = \pi$.

So our new normal ordering is defined as

\begin{eqnarray}
\label{no}
\mbox{{\bf :}} \, {\hat X}^{\mu}(w, \bar w )\, {\hat X}^{\nu}(w^\prime, {\bar w}^\prime)
\,\mbox{{\bf :}} &=&
 {\hat X}^{\mu}(w, \bar w ) \, {\hat X}^{\nu}(w^\prime, {\bar w}^\prime) 
+ \frac{\alpha'}{2}
\eta^{\mu \nu} ln \vert w-w' \vert^2  \nonumber\\
&+& \frac{\alpha'}{2}
\Big( [ \eta + B ]^{-1}\,[ \eta - B ] \Big)^{\mu \nu}\,\, ln (w- \bar{w}')
\nonumber\\ 
&+&  \frac{\alpha'}{2} \Big( [ \eta  + B ] \, [ \eta - B ]^{-1} \,\Big)^{\mu \nu} \,\, 
ln (\bar{w}- w' )\,+\,
\alpha' D^{\mu\nu}\nonumber\\
\end{eqnarray}

\noindent where $D^{\mu\nu}$ is a constant that may depend on $B$ but not 
on the coordinates. 

\section{Equal time commutators}

It is important to investigate the effect of this normal ordering on the 
commutators of position operators to check if the non commutativity of space time
coordinates in the  presence of the antisymmetric tensor background is changed.
We can rewrite eq.(\ref{no}) in a more convenient form for calculating the commutators:

\vskip 1cm
\begin{eqnarray}
\label{n11}
\mbox{{\bf :}} \, {\hat X}^{\mu}(w, \bar w )\, {\hat X}^{\nu}
(w^\prime, {\bar w}^\prime)\,\mbox{{\bf :}} &=&
 {\hat X}^{\mu}(w, \bar w )\, {\hat X}^{\nu}(w^\prime, {\bar w}^\prime) 
+ \frac{\alpha'}{2}
\eta^{\mu\nu}\, ln \vert w-w' \vert^2  \nonumber\\
&-& \alpha' \eta^{\mu\nu}\, ln \vert w-{\bar w}' \vert 
\,+\, 
\alpha' G^{\mu\nu}\,ln \vert w-{\bar w}' \vert^2 \nonumber\\
&+&
{1\over 2\pi} \Theta^{\mu\nu} \, ln \Big({  w-{\bar w}'\over {\bar  w} - w' }\Big)
\,+\,
\alpha' D^{\mu\nu}
\end{eqnarray}

\noindent where we introduced 

\begin{eqnarray}
 G^{\mu\nu} &=& \Big( [ \eta + B ]^{-1}\, \eta \, [\eta - B ]^{-1}\Big)^{\mu\nu}
 \nonumber\\
\Theta^{\mu\nu} &=& -\,2\pi \alpha' \Big( [ \eta + B ]^{-1}\, B \, 
[\eta - B ]^{-1} \Big)^{\mu\nu}
\end{eqnarray}

Now we calculate the normal ordered commutator at boundary points 
$w = {\bar w } = \tau \,\,,\,\, w' = {\bar w}' = \tau'\,$ 
using the same choice for the 
constant $D^{\mu\nu}$ and the same procedure as in \cite{SW}

\begin{eqnarray}
\label{n12}
 \mbox{{\bf :}}\, \Big[ {\hat X}^{\mu} (\tau)\,,\, {\hat X} ^{\nu} 
(\tau' \Big] \mbox{{\bf :}}\,
&\equiv& \mbox{{\bf :}}\, {\hat X}^{\mu}( \tau )\, {\hat X}^{\nu}
( \tau')\,\mbox{{\bf :}}\,-\,
\mbox{{\bf :}}\, {\hat X}^{\nu}( \tau') \, {\hat X}^{\mu} ( \tau )\,\mbox{{\bf :}}
\nonumber\\
&=& \Big[ {\hat X}^{\mu} (\tau)\,,\, {\hat X}^{\nu} (\tau' ) \Big] 
\,+\, \alpha' G^{\mu\nu} ln (( \tau - \tau')^2) 
- {i\over 2}\Theta^{\mu\nu} \epsilon( \tau - \tau' )  
\nonumber\\
&-&  \alpha' G^{\nu\mu} ln (( \tau' - \tau)^2) +
{i\over 2}\Theta^{\nu\mu} \epsilon( \tau' - \tau )  \nonumber\\
&=&  \Big[ {\hat X}^{\mu} (\tau)\,,\, {\hat X}^{\nu} (\tau' ) \Big] 
\end{eqnarray}

So the commutator does not get any extra contribution from the new normal ordering 
prescription. The equal time commutator thus keeps the same form calculated in \cite{SW}
(see also \cite{Braga:2001ci,Banerjee:2002ky}). 

\begin{equation}
\mbox{{\bf :}}\,\Big[ {\hat X}^{\mu} (\tau)\,,\, {\hat X}^{\nu} (\tau ) 
\Big]\,\mbox{{\bf :}}\, \,=\, i \Theta^{\mu\nu}
\end{equation}

Concluding, the new normal ordering for position operators
that is consistent with both equations of motion and boundary conditions at 
quantum level does not spoil the previous results related to 
non commutativity of space time coordinates.

\bigskip
\noindent Acknowledgments: The authors are partially supported by  CNPq., 
FAPERJ and CLAF.

\end{document}